\documentclass[aps,prb,preprint,showpacs,unsortedaddress,superscriptaddress,amsmath,amssymb]{revtex4-1}

\usepackage{bm}
\usepackage{graphicx}
\usepackage{color}
\usepackage{epsfig}
\usepackage{epstopdf}
\usepackage{natbib}
\usepackage{amsmath}

\begin{document}

\title{
\begin{center}
Interplay of spin-orbit torque and thermoelectric effects in ferromagnet/normal metal bilayers
\end{center}}

\author{Can Onur Avci}
\email{can.onur.avci@mat.ethz.ch}
\affiliation{Department of Materials, ETH Z\"{u}rich, H\"{o}nggerbergring 64, CH-8093 Z\"{u}rich, Switzerland}

\author{Kevin Garello}
\affiliation{Department of Materials, ETH Z\"{u}rich, H\"{o}nggerbergring 64, CH-8093 Z\"{u}rich, Switzerland}

\author{Mihai Gabureac}
\affiliation{Department of Materials, ETH Z\"{u}rich, H\"{o}nggerbergring 64, CH-8093 Z\"{u}rich, Switzerland}

\author{Abhijit Ghosh}
\affiliation{Department of Materials, ETH Z\"{u}rich, H\"{o}nggerbergring 64, CH-8093 Z\"{u}rich, Switzerland}

\author{Andreas Fuhrer}
\affiliation{IBM Research - Zurich, S\"{a}umerstrasse 4, CH-8803 R\"{u}schlikon, Switzerland}

\author{Santos F. Alvarado}
\affiliation{Department of Materials, ETH Z\"{u}rich, H\"{o}nggerbergring 64, CH-8093 Z\"{u}rich, Switzerland}

\author{Pietro Gambardella}
\affiliation{Department of Materials, ETH Z\"{u}rich, H\"{o}nggerbergring 64, CH-8093 Z\"{u}rich, Switzerland}

\date{\today}

\begin{abstract}

We present harmonic transverse voltage measurements of current-induced thermoelectric and spin-orbit torque (SOT) effects in ferromagnet/normal metal bilayers, in which thermal gradients produced by Joule heating and SOT coexist and give rise to ac transverse signals with comparable symmetry and magnitude. Based on the symmetry and field-dependence of the transverse resistance, we develop a consistent method to separate thermoelectric and SOT measurements. By addressing first ferromagnet/light metal bilayers with negligible spin-orbit coupling, we show that in-plane current injection induces a vertical thermal gradient whose sign and magnitude are determined by the resistivity difference and stacking order of the magnetic and nonmagnetic layers. We then study ferromagnet/heavy metal bilayers with strong spin-orbit coupling, showing that second harmonic thermoelectric contributions to the transverse voltage may lead to a significant overestimation of the antidamping SOT. We find that thermoelectric effects are very strong in Ta(6nm)/Co(2.5nm) and negligible in Pt(6nm)/Co(2.5nm) bilayers. After including these effects in the analysis of the transverse voltage, we find that the antidamping SOTs in these bilayers, after normalization to the magnetization volume, are comparable to those found in thinner Co layers with perpendicular magnetization, whereas the field-like SOTs are about an order of magnitude smaller.

\end{abstract}

\maketitle 

\section{INTRODUCTION}
\label{Introduction}
Ferromagnet/normal metal (FM/NM) heterostructures host a variety of magnetotransport phenomena that arise from the correlation of electrical, magnetic, and thermal effects. It has been recently shown that an electric current flowing in the plane of a FM/NM bilayer with large spin-orbit coupling generates spin torques that are strong enough to switch the magnetization of the FM.~\cite{MironNM2010,MironN2011,LiuS2012,GarelloNN2013,KimNM2013} These so-called spin-orbit torques (SOTs) have attracted considerable interest as a means to control the magnetic state of spintronic devices~\cite{MironN2011,LiuS2012,HaazenNM2013,CubukcuAPL2014} and motivated extensive investigations into their origin (spin Hall and/or Rashba effect) and dependence on material properties.~\cite{GarelloNN2013,KimNM2013,AvciAPL2012,pai2012APL,FanNC2013,JamaliPRL2013,AvciPRB2014} In these systems, the coupling of charge, heat, and spin currents additionally gives rise to thermoelectric and thermomagnetic phenomena, such as the anomalous Nernst (ANE) and spin Seebeck (SSE) effects.~\cite{BauerNM2012} Both the ANE and SSE have drawn recent attention as they generally coexist and are amplified in strongly spin-orbit coupled FM/NM bilayers.~\cite{UchidaN2008,WeilerPRL2012,HuangPRL2011,KikkawaPRL2013,SchmidPRL2013} NM with large spin-orbit coupling (e.g., Pt) are also commonly used to convert spin into charge currents via the inverse spin Hall effect, since pure spin currents are not directly accessible with electrical measurements.~\cite{UchidaN2008,QuPRL2013}

Because of the strong spin-orbit coupling and the vertical asymmetry inherent to FM/NM stacks, the materials commonly used for SOTs are also suitable for the generation and detection of thermoelectric effects. This may lead to novel strategies to develop functional thermoelectric devices, provided that SOT and thermoelectric phenomena can be correctly identified and measured. In principle, the detection of both SOT and thermoelectric effects is possible within an all-electrical scheme based on harmonic Hall voltage measurements.~\cite{GarelloNN2013} This is a widely employed method to characterize SOTs in FM/NM heterostructures, which is based on measuring the second harmonic changes of the Hall voltage induced by oscillations of the magnetization due to the injection of an ac current.~\cite{GarelloNN2013,KimNM2013,FanNC2013,JamaliPRL2013,AvciPRB2014,PiAPL2010,HayashiPRB2014} Thus far, thermally driven effects in SOT measurements have been reported to be small~\cite{GarelloNN2013,AvciPRB2014} or neglected, while a consistent model and quantitative separation of the SOT and thermoelectric voltage signals has not been attempted. However, the injection of relatively high current densities into ultrathin structures unavoidably causes Joule heating,~\cite{FangohrPRB2011} which can create temperature gradients and consequently generate charge imbalances due to the ANE and SSE. Therefore, SOTs and thermoelectric effects should not be treated independently of each other. This has two implications: First, the signals generated by these effects can add up and lead to ambiguous results for individual measurements of either SOT or thermoelectric properties. Second, current-induced SOTs and thermally-driven spin and charge currents can be intentionally combined to create novel thermoelectric torques.~\cite{HatamiPRL2007}

Motivated by these considerations, we present here a combined study of current driven thermoelectric and SOT effects in different FM/NM bilayers, where FM = Co and NM is either a light metal (LM = Ti, Cu) or a heavy metal (HM = Pt, Ta). The LM and HM pairs are chosen so as to have one element with a much higher resistivity than Co (Ti, Ta), and one element with smaller (Cu) or comparable (Pt) resistivity. By employing harmonic transverse voltage measurements we demonstrate that current injection and consequent Joule heating in FM/LM systems with negligible spin-orbit coupling induces a large second harmonic anomalous Nernst signal due to a vertical thermal gradient, the magnitude and direction of which can be tuned by changing the resistivity or the position of the NM layer relative to the FM layer. We further show how to separate SOT and thermoelectric signals in FM/HM layers where both effects are significant. We find that the thermoelectric transverse voltage contribution is negligibly small in Pt/Co layers, whereas it is considerably larger with respect to the SOT contribution in Ta/Co. The remainder of this paper is organized as follows: Sections~\ref{Setup} and \ref{Harmonic analysis} describe the experimental setup and Harmonic transverse voltage analysis, respectively (see also Appendix~A). This analysis is complemented by macrospin simulations of the transverse voltage (Sect.~\ref{Simulations}) and the separation of SOT and thermal contributions to the second harmonic transverse resistance (Sect.~\ref{Sep_SOT_Thermal}). Finally, the experimental results on FM/LM and FM/HM bilayers are presented in Sect.~\ref{FM-LM results} and \ref{FM-HM results}, respectively.

\section{EXPERIMENTAL DETAILS}
\subsection{Sample preparation and setup}
\label{Setup}
\begin{figure}
  \centering
  \includegraphics[width=8.5 cm]{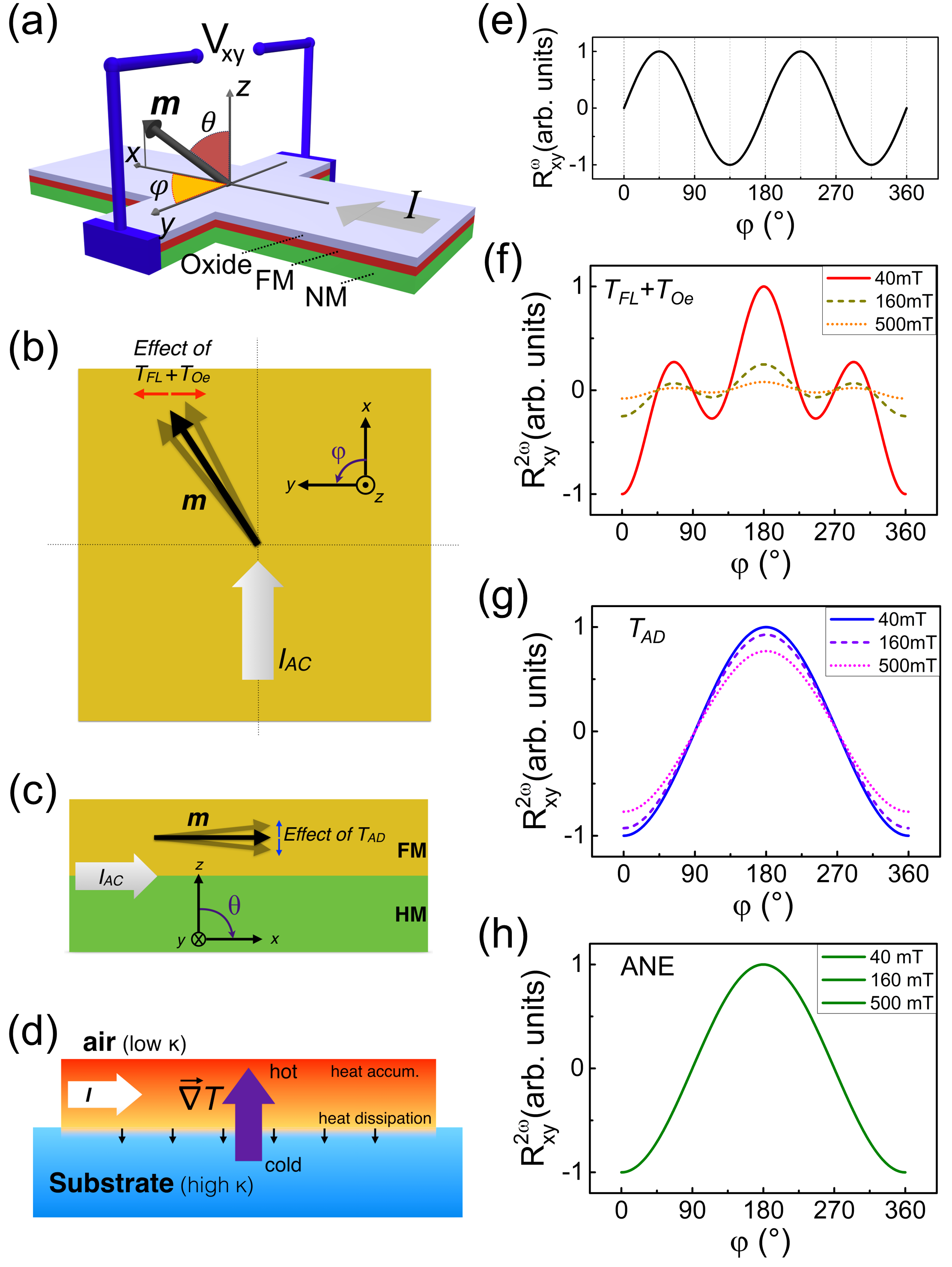}\\
  \caption{(a) Experimental setup and coordinate system. Oscillations of the magnetization due to (b) the field-like SOT and Oersted field ($\mathbf{T}_{FL}+\mathbf{T}_{Oe}$), and (c) antidamping SOT ($\mathbf{T}_{AD}$) induced by an ac current. (d) Schematic of the vertical thermal gradient produced by an in-plane current. Simulations of the (e) first harmonic and (f-h) second harmonic transverse resistance corresponding to (f) field-like torque, (g) antidamping torque, and (h) ANE due to an ac current.}\label{fig1}
\end{figure}

The samples were grown by dc magnetron sputtering on oxidized Si wafers with the following composition: SiO$_2$/NM(6nm)/Co(2.5nm)/Al(1.6nm) and SiO$_2$/Co(8nm)/Al(1.6nm), where NM = Ti, Pt, Ta. Two Cu-based stacks with inverted FM/NM position were also grown, namely SiO$_2$/Ta(1nm)/Cu(6nm)/Co(2.5nm)/Al(1.6nm) and SiO$_2$/Ta(1nm) /Co(2.5nm)/Cu(6nm)/Al(1.6nm), where a 1~nm thick Ta buffer layer was pre-deposited on the SiO$_2$ substrate to induce smooth growth of Cu and Co and enhance the interface quality of the FM. Such a thin Ta layer is a very poor conductor with respect to Cu and Co, and is likely to be oxidized due to large bond enthalpy of TaO (comparable to SiO), so that its influence on the electrical measurements is henceforth neglected. The Al capping layer was oxidized by exposure to an rf O plasma, except in the cases of Pt/Co/Al and Ta/Co/Al, which were oxidized in ambient atmosphere. All samples present isotropic in-plane (easy-plane) anisotropy, with the perpendicular direction being the hard magnetization axis. The as-grown layers were patterned by using standard optical lithography and dry etching methods in the form of Hall bars of width $d$=4 or 5 $\mu$m for the current injection line, $d/2$ for the Hall branches [Fig.~\ref{fig1} (a)], and a separation of $5d$ (not shown on the figure) between two Hall cross regions. The definition of the angles and coordinate system used throughout the paper is given in Fig.~\ref{fig1} (a). For the transverse measurements, the samples were mounted on a motorized stage allowing for in-plane rotation of the angle $\varphi$ and placed in an electromagnet producing fields up to 2~T. All measurements were performed at room temperature with an ac current modulated at $f=10$~Hz.

\subsection{Harmonic transverse resistance measurements}
\label{Harmonic analysis}
It is now established both theoretically and experimentally that an in-plane current flowing in a NM/FM heterostructure with strong spin-orbit coupling generates two qualitatively different types of SOTs:~\cite{GarelloNN2013,KimNM2013,HaneyPRB2013} a field-like (FL) torque $\mathbf{T}_{FL}\sim \mathbf{m}\times \mathbf{y}$, and an antidamping (AD) torque
$\mathbf{T}_{AD}\sim \mathbf{m}\times (\mathbf{y} \times \mathbf{m})$, where $\mathbf{m}$ is the magnetization unit vector and $\mathbf{y}$ is the in-plane axis perpendicular to current flow direction $\mathbf{x}$. When the magnetization lies in the sample plane, the action of $\mathbf{T}_{FL}$ is equivalent to that of an in-plane field $\mathbf{B}_{FL}\sim \mathbf{y}$, and that of $\mathbf{T}_{AD}$ to an out-of-plane field $\mathbf{B}_{AD}\sim \mathbf{m}\times \mathbf{y}$. By injection of a relatively moderate ac current $I=I_{0}\sin(\omega t)$, these fields induce periodic oscillations of the magnetization about its equilibrium position, which is defined by the external, anisotropy, and demagnetizing fields [Fig.~\ref{fig1} (b-c)]. Therefore, the Hall resistance $R_H(t)$ oscillates at a frequency $\omega$ and the Hall voltage $V_H(t)=R_H(t) I_{0}\sin(\omega t)$ has a second harmonic component that relates directly to the current-induced fields.~\cite{PiAPL2010} By defining first and second harmonic Hall resistances, $R^{\omega}_{H}$ and $R^{2\omega}_{H}$, the Hall voltage can be written as $V_H(t)= I_{0}[R_H^{\omega}\sin(\omega t)+R_H^{2\omega}\cos(2\omega t)]$ (see Appendix A). In previous work we have shown that, in addition to the anomalous Hall resistance ($R_{AHE}$), also the planar Hall resistance ($R_{PHE}$) and thermoelectric signals must be taken into account to properly model first and second order effects.~\cite{GarelloNN2013} Here, we consider Joule heating by the injected current as the sole source of a thermal gradient and assume $\nabla T \propto I^{2}R_{s}$, where $R_{s}$ is the sample resistance. For an ac current we thus have

\begin{equation}
\label{eqDT}
\nabla T \propto I_0^{2}\sin^2(\omega t)R_s=\frac{1}{2}I_0^{2}[1-\cos(2\omega t)]R_s \, .
\end{equation}

This relationship implies that the transverse resistance ($R_{xy}(t)$) contains zeroth and second harmonic terms that are proportional to temperature gradients in the sample additional to $R_H$. In metallic FM the most significant thermoelectric voltage driven by a temperature gradient is due to the ANE, which produces an electric field $\mathbf{E}_{ANE}=-\alpha \mathbf{\nabla} T \times \mathbf{m}$, where $\alpha$ is the ANE coefficient. As illustrated in Fig.~\ref{fig1} (d), in-plane current injection through the layers favors the creation of a perpendicular temperature gradient. Due to the large difference of thermal conductivity between the SiO$_2$ substrate ($\kappa$=1.4 Wm$^{-1}$K$^{-1}$) and air ($\kappa$=0.024 Wm$^{-1}$K$^{-1}$) we assume that heat dissipation will take place predominantly via the substrate, inducing a positive thermal gradient in the samples. Inhomogeneous current flow in the top and bottom metal layers can induce an additional contribution to the perpendicular thermal gradient. Note that the geometry that we describe here fulfills also the requirements for the creation and detection of the longitudinal SSE,~\cite{UchidaAPL2010} although the SSE can be expected to be smaller than the ANE in metallic FM/NM bilayers.~\cite{HuangPRL2011,AveryPRL2012,SchmidPRL2013} As the symmetry of the longitudinal SSE signal is the same as that of the ANE signal, our analysis remains valid independently of the microscopic origin of the thermoelectric voltage.


The first and second harmonic expressions for the transverse resistance can finally be written as

\begin{equation}
\label{eqRf}
R^{\omega}_{xy}=R_{AHE}\cos\theta +R_{PHE}\sin^{2}\theta \sin(2\varphi),
\end{equation}
\begin{multline}
\label{eqR2f}
R^{2\omega}_{xy}=[R_{AHE}-2R_{PHE}\cos\theta \sin(2\varphi)]\frac{d\cos\theta}{d\mathbf{B}_{I}}\cdot \mathbf{B}_{I} \\
+R_{PHE}\sin^{2}\theta\frac{d\sin(2\varphi)}{d\mathbf{B}_{I}}\cdot \mathbf{B}_{I}+ I_{0} \alpha \nabla T\sin\theta\cos\varphi,
\end{multline}

where $\theta$ and $\varphi$ are the polar and azimuthal angles of the magnetization vector, respectively, and $\mathbf{B}_{I}=\mathbf{B}_{FL}+\mathbf{B}_{AD}+\mathbf{B}_{Oe}$ represents the sum of the current-induced fields, including the Oersted term, which is assumed to be linearly proportional to the current. Harmonic transverse resistance measurements of SOTs are usually performed as a function of the external magnetic field ($\mathbf{B}_{ext}$) by varying the field magnitude and keeping its direction fixed (field scans).~\cite{GarelloNN2013,KimNM2013,FanNC2013,JamaliPRL2013,AvciPRB2014,PiAPL2010,HayashiPRB2014} For the purpose of this work, however, it is more convenient to consider the case in which $B_{ext}$ is kept constant and its direction changed (angle scans), which we treat in Appendix A for the general situation where $\mathbf{B}_{ext}$ and $\mathbf{m}$ point towards arbitrary directions. If $\mathbf{B}_{ext}$ is applied in-plane and the samples have isotropic in-plane (easy-plane) anisotropy ($\theta=\pi/2$), Eq.~\ref{eqR2f} simplifies to

\begin{equation}
\label{eqR2fsimple}
R^{2\omega}_{xy} = \left [\dfrac{dR^{\omega}_{xy}}{d\theta} \frac{B_{AD}}{B_{ext}}
+\dfrac{dR^{\omega}_{xy}}{d\varphi} \frac{B_{FL}+B_{Oe}}{B_{ext}}+ I_{0} \alpha \nabla T  \right]\,\cos\varphi \, = R^{2\omega}_{AD} + R^{2\omega}_{FL} + R^{2\omega}_{\nabla T}  \, ,
\end{equation}
where the AD, FL (including Oersted field), and thermal contributions to the second harmonic transverse resistance, $R^{2\omega}_{AD}$, $R^{2\omega}_{FL}$, and $R^{2\omega}_{\nabla T}$, appear as separate terms. We note that $R^{\omega}_{xy}$ depends in general on the external field as well as on the effective anisotropy field present in the sample, including the demagnetizing field, as discussed in Sect.~\ref{Sep_SOT_Thermal} in more detail. By carrying out the derivatives of $R^{\omega}_{xy}$ in Eq.~\ref{eqR2fsimple}, we obtain

\begin{equation}
\label{eqR2fsimpleAng}
R^{2\omega}_{xy} = \left[ \left( R_{AHE}\frac{B_{AD}}{B_{ext}}+ I_{0} \alpha \nabla T \right) \cos\varphi
+ 2 R_{PHE}\left( 2\cos^{3}\varphi - \cos\varphi \right) \frac{B_{FL}+B_{Oe}}{B_{ext}}\right] \, .
\end{equation}

Thus $R^{2\omega}_{AD}$ and $R^{2\omega}_{\nabla T}$ are both proportional to $\cos\varphi$ and induce the same angular dependence of $R^{2\omega}_{xy}$, whereas $R^{2\omega}_{FL}$ is proportional to $(2\cos^{3}\varphi - \cos\varphi)$. The above equation shows that, by measuring the dependence of $R^{2\omega}_{xy}$ on the angle $\varphi$, the FL SOT can be separated from the combined contribution of the AD SOT and thermoelectric effects. In Sect.~\ref{Sep_SOT_Thermal} we will show that it is further possible to separate the AD SOT and thermal contributions by measuring the field and current dependence of $R^{2\omega}_{xy}$.




\subsection{Simulations of the transverse signals}
\label{Simulations}
Figure~\ref{fig1} (e-h) shows the simulations of the first and second harmonic transverse resistances corresponding to the equilibrium magnetization and individual action of $\mathbf{B}_{FL}$, $\mathbf{B}_{AD}$ and ANE, respectively. To simulate $R^{\omega}_{xy}$ and $R^{2\omega}_{xy}$, we compute the 
magnetization position by considering the sum of all torques while the external field is rotated in the $xy$ plane between $0^{\circ}$ and $360^{\circ}$. The magnetization is assumed to be uniform while the transverse voltage is calculated using standard expressions for the AHE, PHE, and ANE. The simulations are repeated for positive and negative dc currents for which the half of the difference and the average of these two signals correspond to the equilibrium (current independent) and current induced signals, respectively. This is equivalent to Fourier-transformed first and second harmonic signals in an ac current injection measurements. Note that, relative to the simulations and depending on the system under study, the direction and amplitude of the torques and ANE can change sign in the experiment. The first harmonic signal consists of only the PHE resistance and is proportional to $\sin(2\varphi)$, in agreement with Eq.~\ref{eqRf} when $\theta = 90^{\circ}$. The second harmonic signal shows rather distinct features. As discussed above, $\mathbf{B}_{AD}$ and the ANE are both proportional to $\cos\varphi$ and induce the same angular dependence of $R^{2\omega}_{xy}$, whereas $\mathbf{B}_{FL}$ induces a term proportional to ($2\cos^{3}\varphi - \cos\varphi$), or alternatively [$\frac{1}{2}(\cos3\varphi+cos\varphi)]$, which are both reproduced by the simulations.

\subsection{Separation of FL, AD, and thermal components of $R^{2\omega}_{xy}$}
\label{Sep_SOT_Thermal}
In real measurements, the three signals shown in Fig.~\ref{fig1} (f-h) generally add up and need to be separated into their individual contributions. The simulations show that the $R^{2\omega}_{xy}$ signal due to $\mathbf{B}_{FL}$ vanishes at $\varphi=45^{\circ}$, $135^{\circ}$, $225^{\circ}$, $315^{\circ}$ [Fig.~\ref{fig1} (f)], whereas that due to $\mathbf{B}_{AD}$ and/or the ANE does not vanish [Fig.~\ref{fig1} (g-h)]. A convenient way to separate the $\mathbf{B}_{FL}$ component versus the $\mathbf{B}_{AD}$ plus thermal components is to fit a cosine-like contribution that passes through these four points where the $\mathbf{B}_{FL}$ signal is zero by definition. This fit, which gives $R^{2\omega}_{AD}+R^{2\omega}_{\nabla T}$, contains a combination of $\mathbf{B}_{AD}$ and ANE (and/or SSE), and will be called $\cos\varphi$ contribution in the remainder of the paper. By subtracting $R^{2\omega}_{AD}+R^{2\omega}_{\nabla T}$ from the raw $R^{2\omega}_{xy}$ data one obtains $R^{2\omega}_{FL}$. Notice that all three signals displayed in Fig.~\ref{fig1} (f-h) are symmetric around $\varphi=180^{\circ}$, therefore this separation is valid only if the raw data are also symmetric around $\varphi=180^{\circ}$. Otherwise one needs to find symmetric and antisymmetric parts of $R^{2\omega}_{xy}$ and proceed only with the symmetric part. Antisymmetric signals can occur due to misalignment of the sample with respect to the external field, misalignment of the Hall branches, drift and in-plane temperature gradients due to the fact that the center of the Hall bar is warmer than the contact points. In the measurements presented here the antisymmetric contributions are subtracted from the raw data where applicable. Such antisymmetric effects are found to be of the order of $2-4\%$ of the total signal with the exception of the Co(8nm) sample where it went up to $10\%$ due to enhanced anisotropic thermopower contributions from the in-plane thermal gradient.~\cite{GarelloNN2013}

Further separation of $R^{2\omega}_{AD}$ and $R^{2\omega}_{\nabla T}$ is possible by performing measurements as a function of the external field. The contribution of $\mathbf{B}_{AD}$ to $R^{2\omega}_{xy}$ is a dynamic effect resulting from the oscillations of the magnetization. Thermal contributions, on the other hand, result from a static effect and enter into $R^{2\omega}_{xy}$ through the second order dependence of $\nabla T$ on $I^2$ (Eq.~\ref{eqDT}). Thus the SOT contribution tends to vanish as $B_{ext}$ is large enough to force the magnetization to align rigidly along the field direction, that is, when the susceptibility of the magnetization to an applied field goes to zero. The ANE and SSE, on the other hand, depend only on the magnetization direction and are independent of the external field amplitude (provided that the magnetization is saturated). In order to exploit this difference, we notice that the FL and AD terms in Eq.~\ref{eqR2fsimple} are proportional to the inverse of the external field times the derivative of $R^{\omega}_{xy}$ with respect to, respectively, $\varphi$ and $\theta$. Since we assume negligible in-plane anisotropy and the magnetization is saturated in-plane the PHE is independent of $B_{ext}$ and $\left.\frac{dR^{\omega}_{xy}}{d\varphi}\right|_{B_{ext}} \approx \mathrm{constant}$. Hence $R^{2\omega}_{FL}$ will be inversely proportional to $B_{ext}$ (Eq.~\ref{eqR2fsimple}) and $B_{FL}$ independent of $B_{ext}$ (Eq.~\ref{eqbphi}). On the other hand, the derivative of $R^{\omega}_{xy}$ with respect to $\theta$ near $\theta=90^{\circ}$ depends on the AHE and therefore on the out-of-plane tilt of the magnetization. During the measurement of $\left.\frac{dR^{\omega}_{xy}}{d\theta}\right|_{B_{ext}}$ in the vicinity of $\theta=90^{\circ}$, the out-of-plane component of the external field increases linearly with $B_{ext}$. The action of this component, however, is counteracted by the in-plane component of the external field ($\approx B_{ext}$), the demagnetizing field $B_{dem}\sim\mu_{0}M_{s}$, and sample-dependent perpendicular anisotropy field $B_{ani}=\frac{2K}{M_s}$, where $M_s$ is the saturation magnetization and $K$ is the uniaxial anisotropy constant. We note that, although the magnetization lies in-plane in the absence of an external field, there can be a perpendicular anisotropy field due to interface contributions whose action is opposed to that of the demagnetizing field. This will effectively reduce the field required to saturate the magnetization out of plane, which by definition is ($B_{dem}-B_{ani}$). As a result, we have that $\left.\frac{dR^{\omega}_{xy}}{d\theta}\right|_{B_{ext}}\sim\frac{B_{ext}}{B_{ext}+B_{dem}-B_{ani}}$.
Summarizing these considerations, we find the following qualitative relationships between the second harmonic transverse resistance components and the static fields acting on magnetization:

\begin{equation}
\label{R2f prop to}
\begin{array}{ccc}
R^{2\omega}_{FL}\sim\dfrac{1}{B_{ext}} \, , \quad R^{2\omega}_{AD}\sim\dfrac{1}{B_{ext}+B_{dem}-B_{ani}} \, , \quad R^{2\omega}_{\nabla T}\sim \mathrm{constant} \,.
\end{array}
\end{equation}

These relationships, which have been additionally validated by macrospin simulations, indicate an effective way of separating the transverse resistance contributions due to dynamic (SOT) and static (thermal) effects. Accordingly, the AD and FL components of the current-induced field can then be calculated as:

\begin{eqnarray}
\label{eqbtheta}
B_{AD} & = & \left[R^{2\omega}_{AD}/\left(\cos\varphi\dfrac{dR^{\omega}_{xy}}{d\theta}\right)\right]B_{ext} \\
\label{eqbphi}
B_{FL} + B_{Oe} & = & \left[R^{2\omega}_{FL} /\left(\cos\varphi\dfrac{dR^{\omega}_{xy}}{d\varphi}\right)\right]B_{ext} \, .
\end{eqnarray}

\section{RESULTS}
\label{results}

\subsection{Thermoelectric effects in FM/LM layers}
\label{FM-LM results}

\begin{figure*}
  \centering
  \includegraphics[width=16 cm]{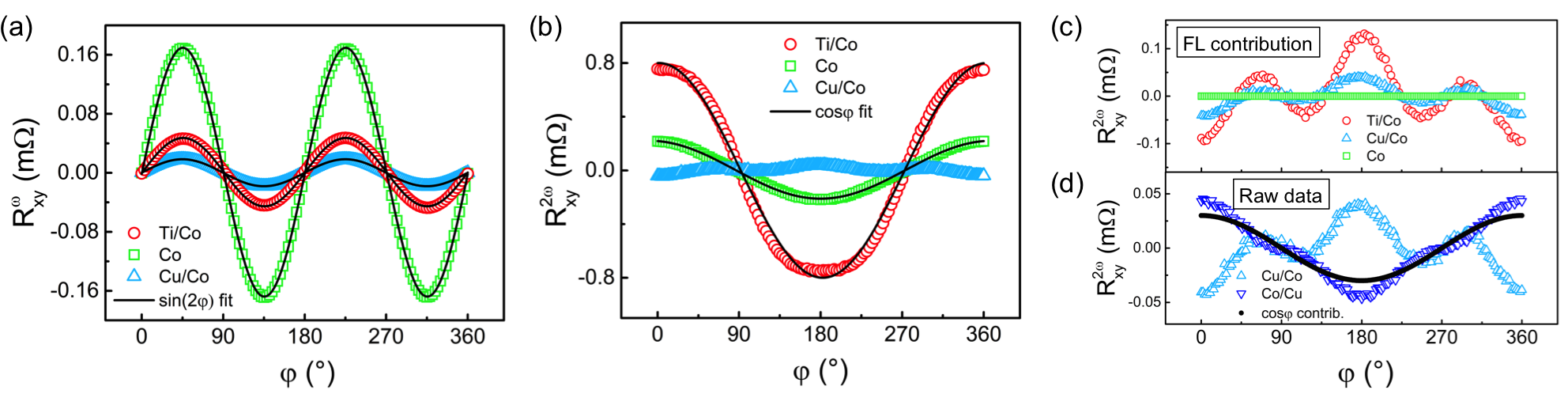}\\
  \caption{(a) $R^{\omega}_{H}$ and (b) $R^{2\omega}_{H}$ measured for Ti/Co, Cu/Co and Co layers. (c) $R^{2\omega}_{FL}$ component of the second harmonic signal obtained by subtraction of the $\cos\varphi$ fits performed by taking into account the symmetry considerations given in Sect.~\ref{Sep_SOT_Thermal}. (d) Comparison of $R^{2\omega}_{H}$ for Cu/Co and Co/Cu inverted stacks. All the measurements are performed at $j=10^7$A/cm$^2$ and $B_{ext}=200$~mT, except for the inverted Co/Cu sample for which $B_{ext}=80$ mT in order to show data with comparable Oersted and ANE contributions. A small constant offset due to misalignment of the Hall branches are subtracted from all first and second harmonic signals.}\label{fig2}
\end{figure*}

In order to verify our hypothesis on the generation and detection of thermal effects, we have performed transverse resistance measurements on Ti/Co and Cu/Co layers, as well as on the reference Co and inverted Co/Cu layers. These LM were specifically chosen so as to minimize any spin-orbit coupling effect and to compare the role played by the resistivity and position of the LM relative to the FM layer. The resistivity is expected to be at least one order of magnitude higher in Ti with respect to Cu considering their bulk values, while the resistivity of Co is in between the two. We have injected an ac current of 4 mA (Co), 4.25 mA (Ti/Co, Cu/Co) and 3.4 mA (Co/Cu), equivalent to a current density of 10$^{7}$ A/cm$^{2}$ (differences are due to variations in the device size), and measured the transverse resistance with the external field set to 200 mT and rotated in the $xy$ plane in steps of $2^{\circ}$. Figure~\ref{fig2} (a) shows $R^{\omega}_{xy}$, which has the typical $\sin(2\varphi)$ dependence expected of $R_{PHE}$ (Eq.~\ref{eqRf} for $\theta = 90^{\circ}$). Sinusoidal fits (solid curves) show that the magnetization strictly follows the external field, indicating that the in-plane magnetic anisotropy is negligibly small. Due to the current flow in the NM, which does not contribute to the transverse voltage, $R^{\omega}_{xy}$ is lower in Cu/Co and Ti/Co layers relative to Co. The resistivities of the samples, measured using a four point geometry, are 34.9~$\mu\Omega$cm for Co, 176.5~$\mu\Omega$cm for Ti/Co, 17.4~$\mu\Omega$cm for Cu/Co, and 14.5~$\mu\Omega$cm for Co/Cu (assuming no current flow in the 1~nm thick Ta seed layer), confirming that Cu is the most and Ti the least conductive layer. By combining the transverse and longitudinal resistivity measurements we conclude that the current is shunted mostly towards the Cu side in Cu/Co and towards the Co side in Ti/Co.

Figure.~\ref{fig2} (b) shows $R^{2\omega}_{xy}$ measured simultaneously with $R^{\omega}_{xy}$. Distinct behaviors are observed for all three samples. In the Ti/Co and Co layers, we recognize a dominant $\cos\varphi$ component, as expected from either the ANE due to a vertical temperature gradient or $\mathbf{B}_{AD}$, according to the simulations reported in Fig.~\ref{fig1} (g-h). The cosine fit (solid lines) matches accurately the Co data, whereas a slight deviation is observed for the Ti/Co bilayer. In Cu/Co, on the other hand, the signal with $\cos\varphi$ symmetry is absent but there is a clear signal with $\mathbf{B}_{FL}$ symmetry $(\cos3\varphi+\cos\varphi)$, as shown by Eqs.~\ref{eqR2fsimple} and \ref{eqR2fangle} as well as by the simulations in Fig.~\ref{fig1} (f).

By using the procedure outlined in the previous section we have separated the cosine ($R^{2\omega}_{AD}+R^{2\omega}_{\nabla T}$) and the FL ($R^{2\omega}_{FL}$) contributions in all three samples. $R^{2\omega}_{FL}$ is shown in Fig.~\ref{fig2} (c), where we observe that $R^{2\omega}_{FL}=R^{2\omega}_{xy}$ for Cu/Co and $R^{2\omega}_{FL}=0$ for Co. These signals are compatible with the symmetry and direction of the Oersted field [see simulation in Fig.~\ref{fig1} (f)] due to the current flow in the NM layer. Using Eq.~\ref{eqbphi} we find $B_{Oe}=-0.26\pm0.19$~mT for Ti/Co, $B_{Oe}=-0.22\pm0.06$~mT for Cu/Co, $B_{Oe}=+0.18\pm0.02$~mT for Co/Cu, and $B_{Oe}\approx0$~mT for Co. We note that a homogeneous current distribution in the NM/FM bilayers would give $B_{Oe}=-0.36$~mT. In Ti/Co layers the measured field is lower than the estimated value which is somewhat expected due to current shunting towards the Co side. However in both Cu/Co and Co/Cu layers we have found values below the estimated one whereas the opposite is expected. We have no explanation for this issue, which may be due to errors in the thickness calibration of the Cu layers. Nonetheless, the comparison of $R^{2\omega}_{xy}$ for the Cu/Co and Co/Cu inverted bilayers, shown in Fig~\ref{fig2} (d), reveals a change of sign consistent with that expected from the Oersted field. Further, on top of the Oersted field contribution of the Co/Cu sample we recognize an additional $cos\varphi$ contribution (solid curve). This signal is constant as a function of the external field, which identifies it as a thermoelectric effect.  Note that we do not expect any contribution to SOT and thermal effects from the 1~nm thick Ta buffer layer: first, because of its likelihood to be oxidized (as mentioned in Sect.II A) and, second, because of the difference in thickness (1:6) and resistivity ($\sim$1:10) between Ta and Cu, which implies that the current distribution in the two layers would scale approximately as 1:60 assuming a fully metallic Ta buffer. Contributions to the $R^{2\omega}_{xy}$ signal reported in Fig~\ref{fig2} (d) due to electrical conduction in Ta can thus be safely neglected. We conclude that bilayers with nominally the same composition and similar resistivity exhibit different thermoelectric responses by just altering the stacking order.

\begin{figure*}
  \centering
  \includegraphics[width=16 cm]{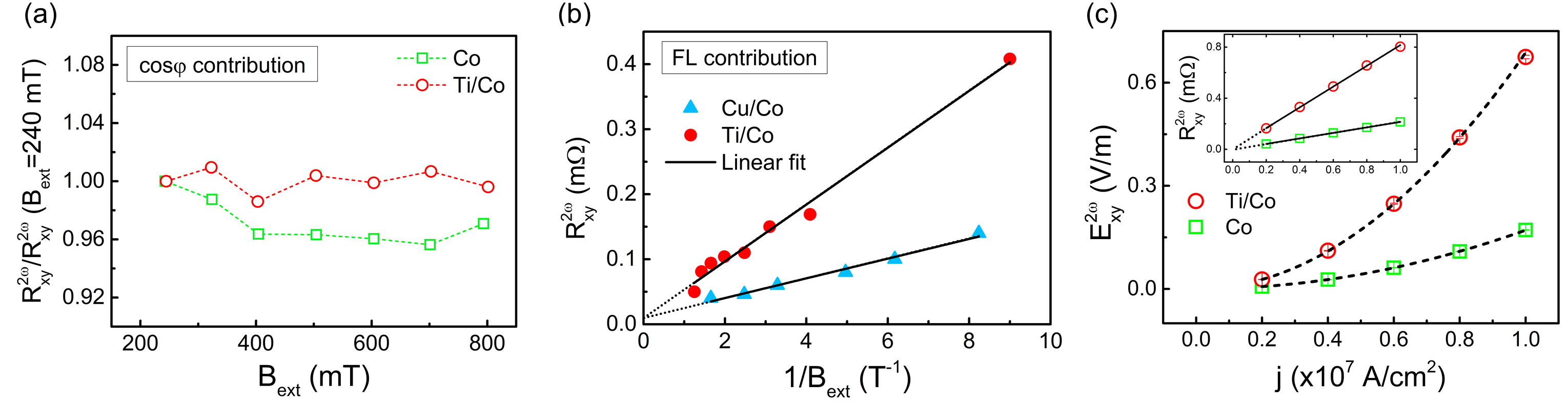}\\
  \caption{(a) $\cos\varphi$ contribution of $R^{2\omega}_{H}$ normalized to the value recorded at $B_{ext}=240$~mT as a function of the external field for Ti/Co and Co. (b) $R^{2\omega}_{FL}$ as a function of the inverse external field. (c) $\cos\varphi$ signal amplitudes (electric field in the main panel, resistance in inset) as a function of the injected current density.}\label{fig3}
\end{figure*}

To further investigate the origin of the different $R^{2\omega}_{xy}$ components, especially the $\cos\varphi$ contributions, we have performed measurements at different external field values. Figure~\ref{fig3} (a) and (b) show the field dependence of the second harmonic transverse resistance amplitudes (the difference between maximum and minimum) after separation of the cosine and FL contributions. For comparison, the signals with $\cos\varphi$ symmetry of the Co and Ti/Co samples in (a) have been normalized to their respective values recorded at $B_{ext}=240$~mT. These signals are unaffected by the external field within an accuracy of 5\%, confirming that the AD-SOT is negligible in these samples and that the $\cos\varphi$ contributions originates from the ANE. This is not surprising since in a single Co layer there is no known mechanism that can give rise to SOT, and Ti is a LM with weak spin-orbit coupling. The signal with FL symmetry is shown in Fig.~\ref{fig3} (b) as a function of the inverse of the external field. The data are proportional to $1/B_{ext}$, as expected from Eq.~\ref{R2f prop to}, and converge to zero as 1/$B_{ext}\rightarrow 0$. This further confirms the Oersted field origin of the FL signal.

To establish the sign of the temperature gradient in the Co reference layer, we consider the equivalent action of a dc current injected along $\pm \mathbf{x}$. According to Fig.~\ref{fig2} (b), when the magnetization is along +$\mathbf{x}$ we measure a positive second harmonic signal, meaning that for positive (negative) current direction $\mathbf{E}_{ANE}$ increases (decreases) the Hall voltage. This indicates that, in our measurement geometry, $\mathbf{E}_{ANE}$ points along -$\mathbf{y}$ for a positive current. As a result, by taking the ANE coefficient $\alpha$ to be positive and considering $\mathbf{m}$ pointing towards +$\mathbf{x}$, we find the temperature gradient to be along +$\mathbf{z}$ direction, consistently with our expectations.

In order to compare the ANE in different samples, we compare the electric fields induced by the thermal gradient, $E^{2\omega}_{\nabla T} = R^{2\omega}_{\nabla T}I_{0}/d$. Figure~\ref{fig3} (c) shows the amplitude of the $\cos\varphi$ contribution of the measured second harmonic electric field, $(R^{2\omega}_{AD}+R^{2\omega}_{\nabla T})I_{0}/d\approx R^{2\omega}_{\nabla T}I_{0}/d$, as a function of the applied current density $j$ in Ti/Co and Co layers. Fits to the data (dashed lines) show that the electric field scales with the square of the injected current density, or, equivalently, that the measured transverse resistance scales linearly with the current (inset), as expected for $R^{2\omega}_{\nabla T}$ (see Eq.~\ref{eqR2fsimple}). We find $E^{2\omega}_{\nabla T} = 0.68$~V/m for Ti/Co and 0.16~V/m for Co layers for $j=10^{7}$ A/cm$^{2}$. Assuming the same ANE coefficients in both layers, this large difference could be explained by much larger resistance of Ti/Co relative to Co and assuming that $\nabla T \sim T \sim I^2R_s$. However, this argument fails for the case of Cu/Co, for which we would expect approximately half of the thermal signal of the Co reference layer and instead we find a negligible $\cos\varphi$ contribution [Fig.~\ref{fig2} (b)]. Moreover, the same argument does not explain why a small thermal signal is detected by inverting the position of the Cu and Co layers, as shown in Fig.~\ref{fig2} (d).

In order to explain this discrepancy we must consider the current distribution inside the bilayer, where the current preferably flows through the more conductive layer, together with the asymmetric heat dissipation towards the air and substrate side of the samples. If we consider a simple model where each layer is represented by an individual resistance ($R_{NM},R_{FM}$), the current flow will be inversely proportional to the resistance since $R_{NM} I_{NM}=R_{FM} I_{FM}$. However, as Joule heating scales with the inverse of the resistance, the less resistive layer will heat more than the more resistive one. This leads to a positive (negative) thermal gradient if the less (more) resistive layer is placed on top, viceversa if it is placed on the bottom. Adding the effect of heat dissipation to such a model leads to an enhancement (decrease) of the thermal gradient when the less resistive layer is placed on top (bottom), because the thermal conductivity of air is much smaller compared to that of the substrate. Accordingly, the heat produced by current flow in the Cu layer of Cu/Co dissipates directly into the substrate producing a negligible thermal gradient in the top Co layer, whereas we observe a positive thermal gradient in the inverted Co/Cu bilayer. The same model explains why thermal effects are enhanced when a strongly resistive NM layer such as Ti (and Ta, see Sect.~\ref{FM-HM results_T}) is placed between a less resistive FM and the substrate. In order to estimate the temperature gradient in our layers we assume an average normalized ANE coefficient of $2.1\times10^{-7}$~VK$^{-1}$T$^{-1}$ within the ones listed in Ref.~\onlinecite{WeischenbergPRB2013} for Co films with [001] texture. Although our layers are polycrystalline, we use this value multiplied by the saturation magnetization of our samples (1.45~T) to obtain an estimate of the ANE coefficient $\alpha=0.31$~$\mu$VK$^{-1}$. By assuming a linear temperature gradient, we find a temperature difference between the top and bottom Co interfaces of 5.57~mK in Ti/Co, 4.46~mK in Co, and 0.21~mK in Co/Cu. Scaling Joule heating as $j^2$ for the different current density, these values appear to be reasonable when compared to other measurements of nm-thick FM/NM bilayers.~\cite{SlachterNP2010,FangohrPRB2011,SlachterPRB2011,SchreierAPL2013}

\subsection{SOT and thermoelectric effects in FM/HM layers}
\label{FM-HM results}
\begin{figure}
  \centering
  \includegraphics[width=8.5 cm]{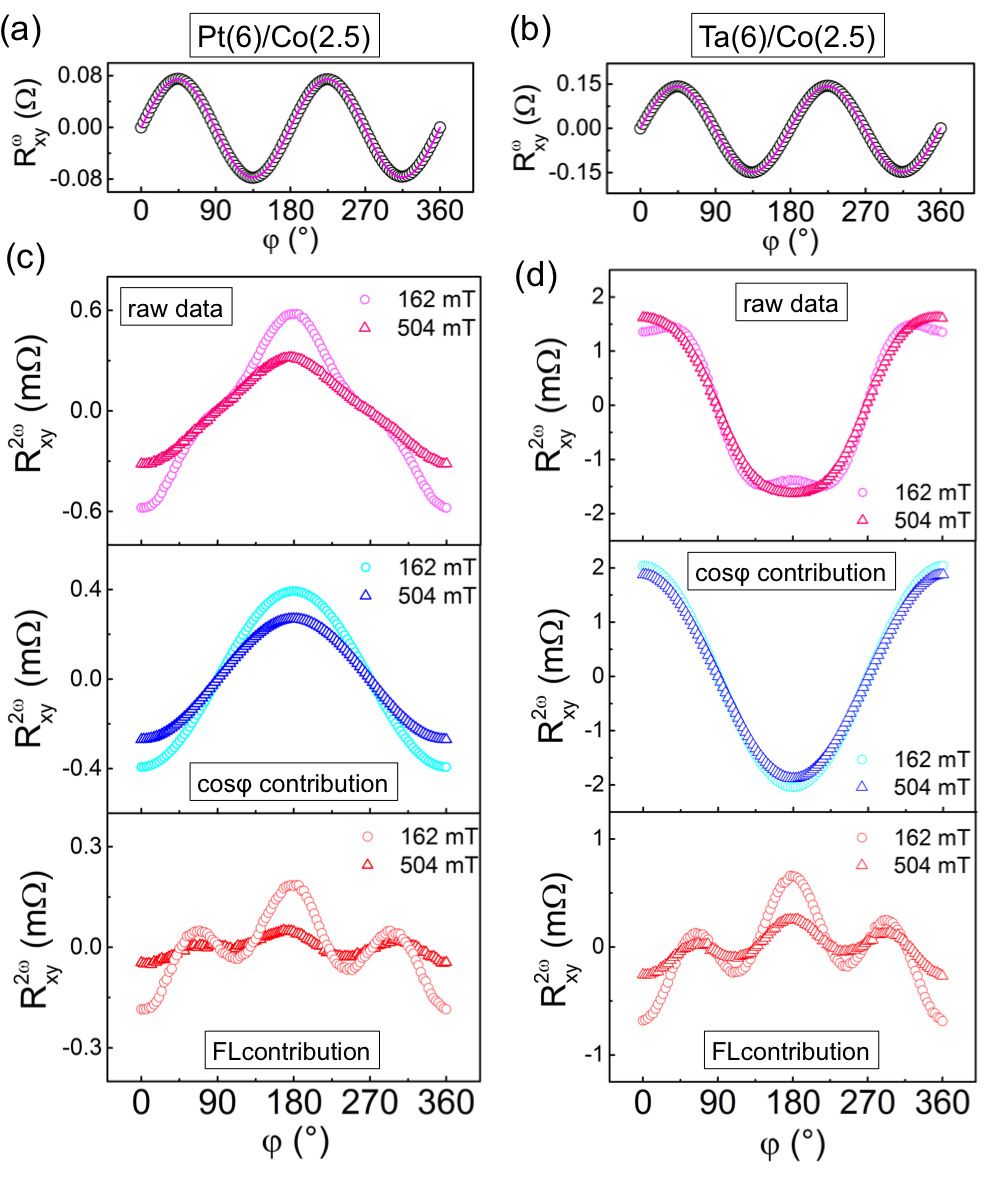}\\
  \caption{(a) $R^{\omega}_{xy}$ of Pt/Co and (b) Ta/Co measured at 162~mT. The solid line is a $\sin(2\varphi)$ fit of the experimental data. (c,d) Top panels: $R^{2\omega}_{xy}$ of Pt/Co (c) and Ta/Co (d) for 2 different applied fields. Middle panels: $R^{2\omega}_{AD}+R^{2\omega}_{\nabla T}$. Bottom panels: $R^{2\omega}_{FL}$. A small constant offset due to misalignment of the Hall branches has been subtracted from the $R^{\omega}_{xy}$ and $R^{2\omega}_{xy}$ signals.}\label{fig4}
\end{figure}

We consider now Pt/Co and Ta/Co bilayers where spin-orbit coupling is strong. Figure~\ref{fig4} (a) and (b) show $R^{\omega}_{xy}$ of Pt/Co and Ta/Co, respectively, measured by rotating the sample in the $xy$ plane in a fixed external field of 162~mT (black open circles). Fits to the data according to Eq.~\ref{eqRf} for $\theta = 90^{\circ}$ are shown as solid curves. We note that $R^{\omega}_{xy}$ measured at higher field does not change, whereas $R^{\omega}_{xy}$ decreases when $B_{ext} \leq 100$~mT due to the unsaturated magnetization. $R^{2\omega}_{xy}$, on the other hand, has a significant field dependence in both bilayers, as shown in the top panels of Fig.~\ref{fig4} (c) and (d). At relatively low field (162 mT), where we expect a higher susceptibility of the magnetization to the SOTs, $R^{2\omega}_{xy}$ has a complex behaviour as a function of $\varphi$, whereas at relatively high field (504~mT) $R^{2\omega}_{xy}$ converges to a $\cos\varphi$ signal. The middle and bottom panels of Fig.~\ref{fig4} (c) and (d) show the separation of the second harmonic signal into the $\cos\varphi$ contribution ($R^{2\omega}_{AD}+R^{2\omega}_{\nabla T}$) and FL contribution ($R^{2\omega}_{FL}$). We observe that $R^{2\omega}_{AD}+R^{2\omega}_{\nabla T}$ changes sign in Ta/Co with respect to Pt/Co. Moreover, this signal has a weak field dependence in Ta/Co and a relatively stronger field dependence in Pt/Co. On the other hand, $R^{2\omega}_{FL}$ has the same sign and similar behavior as a function of the external field in both systems.

To further examine and compare the field dependence of the second harmonic signals we plot the amplitude of $R^{2\omega}_{AD}+R^{2\omega}_{\nabla T}$ and $R^{2\omega}_{FL}$ as a function of the external field, normalized to 1 at $B_{ext}=162$~mT [Fig.~\ref{fig5} (a) and (b)]. We do not choose a lower external field value for the normalization since the magnetization must be completely saturated in both samples. We observe that $R^{2\omega}_{AD}+R^{2\omega}_{\nabla T}$ decreases very fast with increasing $B_{ext}$ in Pt/Co, slower in Ta/Co, and slowest in the reference Co layer. The signal for Co is solely due to the ANE and serves for comparison. We attribute the difference between Pt/Co and Ta/Co to the existence of a significant thermoelectric effect in Ta/Co, which produces a constant $R^{2\omega}_{\nabla T}$ term that offsets the field dependence of the AD-SOT term. Contrary to the cosine-type $R^{2\omega}_{AD}+R^{2\omega}_{\nabla T}$ contribution, $R^{2\omega}_{FL}$ behaves similarly in both systems, showing a fast decrease and approaching to values nearly zero as the field is increased (the FL term is absent in the Co reference layer and thus not plotted).

\subsubsection{Thermoelectric effects in FM/HM layers}
\label{FM-HM results_T}
\begin{figure*}
  \centering
  \includegraphics[width=14 cm]{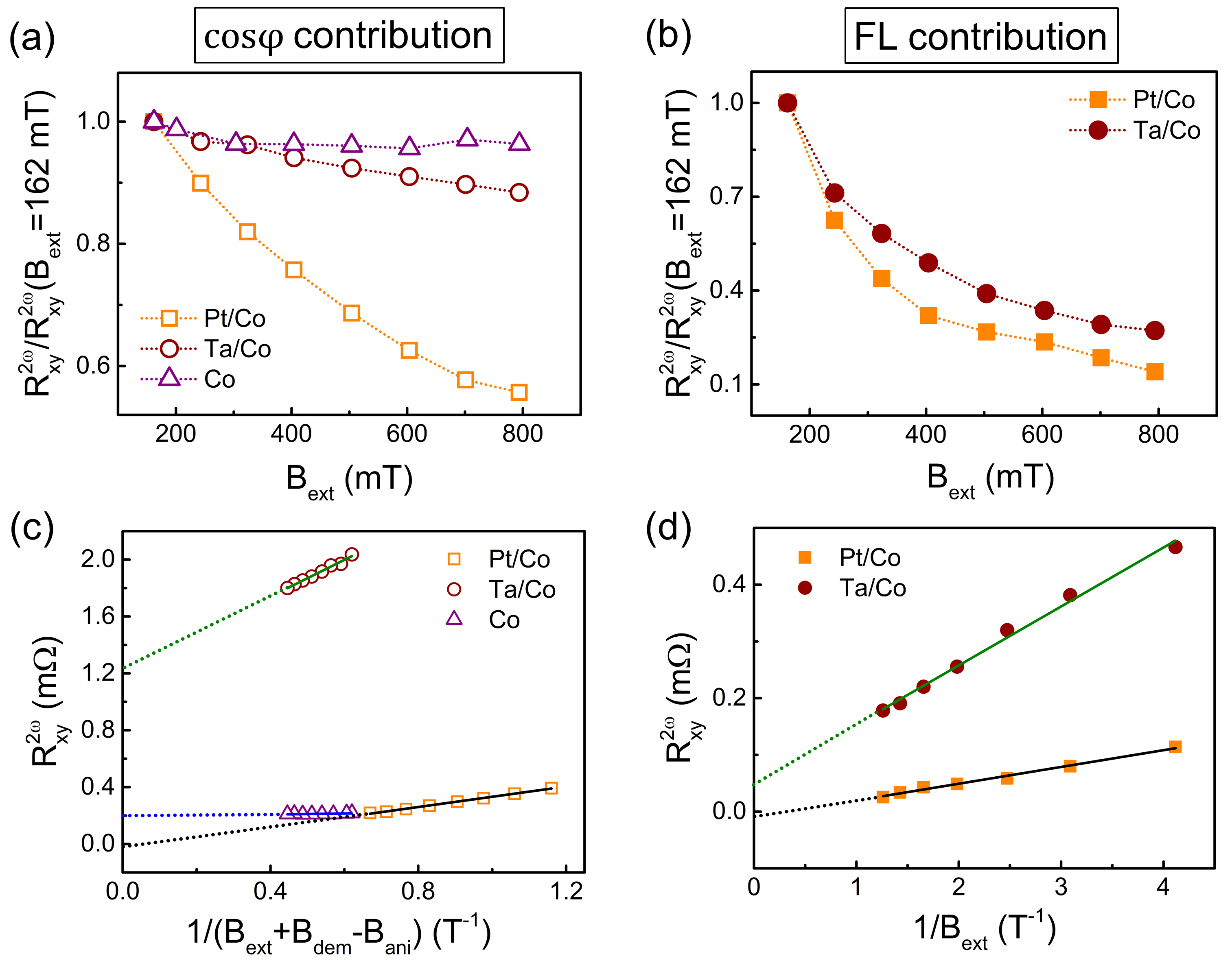}\\
  \caption{External field dependence of (a) $R^{2\omega}_{AD}+R^{2\omega}_{\nabla T}$ and (b) $R^{2\omega}_{FL}$ in Pt/Co, Ta/Co and Co normalized to the values at $B_{ext}=162$~mT. (c) $R^{2\omega}_{AD}+R^{2\omega}_{\nabla T}$ and (d) $R^{2\omega}_{FL}$ as a function of the inverse of the static fields acting against the current-induced field in each case (see Eq.~\ref{R2f prop to}).}\label{fig5}
\end{figure*}

In order to quantitatively separate the thermal and AD-SOT contributions to the $\cos\varphi$-like component of $R^{2\omega}_{xy}$, we exploit the different field dependence of SOT and thermoelectric effects (see Sect.~\ref{Sep_SOT_Thermal}). Figure~\ref{fig5} (c) shows that $R^{2\omega}_{AD}+R^{2\omega}_{\nabla T}$ is a linear function of $\frac{1}{B_{ext}+B_{dem}-B_{ani}}$, as expected from Eq.~\ref{R2f prop to}. Here we have taken $B_{dem}=1.45$~T for all layers, $B_{ani}=0.65$~T for Pt/Co and $B_{ani}=0$~T for Ta/Co. These values were determined by measuring the field required to saturate the magnetization out-of-plane, which is 1.45~T for both the Co and Ta/Co layers, and 0.8~T for Pt/Co. This indicates that the perpendicular magnetic anisotropy is significantly larger for Pt/Co ($\approx$0.65~T) compared to Ta/Co ($\approx$0~T). Linear fits to the data reveal that Ta/Co has a constant offset of 1.22~m$\Omega$, which we associate to thermoelectric effects, whereas the Pt/Co data converge to zero at high field. The data from the Co reference layer are nearly constant and converge to 0.2~m$\Omega$ in the high field limit. Figure~\ref{fig5} (d) shows that $R^{2\omega}_{FL}$ also obeys Eq.~\ref{R2f prop to}, being proportional to $\frac{1}{B_{ext}}$. Both the Pt/Co and Ta/Co data converge towards values near zero (the small residual offset for Ta/Co represent $\approx 1.5\%$ of the raw data and depends on the accuracy of the magnetization angle as well as possible unintentional misalignment of $B_{ext}$ with respect to the $xy$ plane).

This analysis confirms that there is a significant thermoelectric effect in Ta/Co that adds to the AD-SOT second harmonic signal, which is not found for Pt/Co. To separate thermal and AD-SOT effects, we take $R^{2\omega}_{\nabla T}$ equal to the $y$-axis intercept of the linear fit in Fig.~\ref{fig5} (c). We thus obtain $E^{2\omega}_{\nabla T}=1.06$~V/m, a value higher than the one found for Ti/Co ($E^{2\omega}_{\nabla T}=0.68$~V/m). As the resistivity of the Ta/Co sample (142.9~$\mu\Omega$cm) is about 20$\%$ lower compared to Ti/Co (176.5~$\mu\Omega$cm), we would expect a smaller thermal gradient for Ta/Co and thus a reduced ANE relative to Ti/Co. However, the presence of the HM interface may effectively alter the ANE coefficient in FM/HM layers, enhancing it in Ta/Co relative to Ti/Co.

In order to shed light on the $\it{absence}$ of the ANE signal in Pt(6nm)/Co(2.5nm) bilayers we have performed experiments with thinner Pt layers, namely Pt(1-3nm)/Co(2.5nm). As the Pt resistivity increases with decreasing thickness, the current distribution within the Pt/Co bilayer changes. We have observed that the resistivity of Pt/Co increased from 40.3 up to 66.4~$\mu\Omega$cm while decreasing the Pt thickness from 6 to 1~nm. A nonzero thermoelectric signal in agreement with the sign of the ANE was observed as the Pt thickness was $\leq2$~nm. We have found $E^{2\omega}_{\nabla T}=0.21$~V/m for Pt(1nm)/Co and $E^{2\omega}_{\nabla T}=0.08$~V/m for Pt(2nm)/Co layers. These results suggest that the decrease of the signal in thick Pt samples is due to current shunting towards Pt side.

In the FM/HM layers, a vertical temperature gradient can give rise to the SSE in addition to the ANE, leading to an enhanced or decreased $R^{2\omega}_{\nabla T}$ depending on the relative sign of the two effects. In order to verify this point we have performed harmonic Hall measurements on Pt(6nm)/yttrium iron garnet(50nm) samples grown on gadolinium gallium garnet by pulsed laser deposition and sputtering, respectively. Our measurements are the AC equivalent of the ones reported in Ref.~\onlinecite{SchreierAPL2013}. The SSE manifests itself in the second harmonic signal in the same way as the ANE. By properly taking into account the position of the HM with respect to the FM layer and the sign of the spin Hall angle in each system, we find that the SSE, if present, should have the same sign in Pt/Co, and opposite sign in Ta/Co, with respect to the ANE signal. This indicates that neither the signal enhancement in Ta/Co nor the reduction in Pt/Co with respect to expectations can be explained by the action of the SSE.

\begin{figure}
  \centering
  \includegraphics[width=8.5 cm]{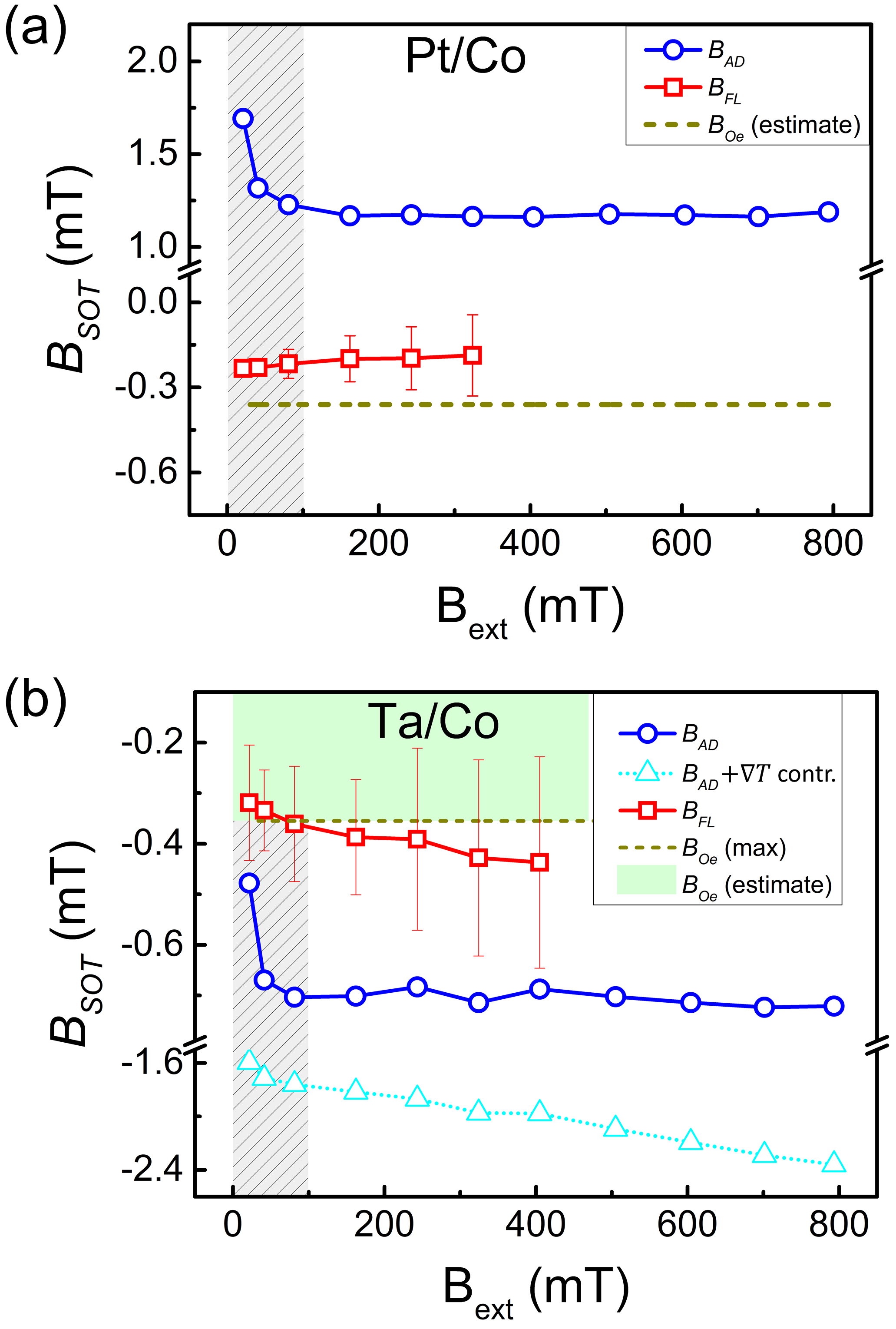}\\
  \caption{FL and AD torques as a function of the external field in (a) Pt/Co and (b) Ta/Co bilayers. The current density is $j=10^7$~A/cm$^{2}$ in both samples. The hatched gray area encloses unreliable data due to incomplete saturation of the magnetization. The shaded green area in (b) shows the range of the Oersted field in Ta/Co, depending on the current distribution within the bilayer.}\label{fig6}
\end{figure}

\subsubsection{SOT in FM/HM layers}
\label{FM-HM results_SOT}
To find the second harmonic signal solely due to the AD-SOT, we subtract $R^{2\omega}_{\nabla T}$ determined above from the total $(R^{2\omega}_{\nabla T}+R^{2\omega}_{AD})\sim \cos\varphi$ signal. The separation of the $R^{2\omega}_{AD}$ and $R^{2\omega}_{FL}$ terms thus allows us to determine the SOT fields using Eqs.~\ref{eqbtheta} and \ref{eqbphi}. The derivative $\dfrac{dR^{\omega}_{H}}{d\varphi}$ appearing in Eq.~\ref{eqbphi} is readily calculated from the curves shown in Fig.~\ref{fig5} (a,b). The derivative $\dfrac{dR^{\omega}_{H}}{d\theta}$ appearing in Eq.~\ref{eqbtheta}, however, is not accessible by angular scans in the $xy$ plane. We thus performed additional measurements of $R^{\omega}_{xy}$ while rotating the external field between $\theta=80^{\circ}$ and $100^{\circ}$, repeating the measurement at each external field value and computed the derivative accordingly. The SOT fields of Pt/Co and Ta/Co are plotted in Fig.~\ref{fig6} (a) and (b). Blue open circles and red open squares represent the effective fields $B_{AD}$ and $(B_{FL}+B_{Oe})$ corresponding to $\mathbf{T}_{AD}$ and $\mathbf{T}_{FL}+\mathbf{T}_{Oe}$, respectively. Hatched areas in gray show unreliable data due to the unsaturated sample magnetization. Below 100~mT, the $R_{PHE}$ value undergoes a relatively sharp decrease for both Pt/Co and Ta/Co layers indicating a non-uniform magnetization in the sample. Above this threshold field value, the variations in the $R_{PHE}$ is negligibly small and the macrospin assumption is valid. For a more accurate interpretation of the data we have drawn the estimated Oersted field assuming homogeneous current flow within the bilayer (dashed line). In Ta/Co, similar to Ti/Co, the injected current is likely to shunt towards the Co side and yield a smaller Oersted field. Therefore its contribution is estimated to be somewhere above the dashed line (green shaded area). Within the error of the measurements, we find that $B_{AD}$ and $B_{FL}$ do not depend on $B_{ext}$, as expected. However, since $R^{2\omega}_{FL}$ decreases rapidly as a function of the external field [Fig.~\ref{fig5} (b)] the signal-to-noise ratio decreases and it is not possible to extend the quantification of $(B_{FL}+B_{Oe})$ to the entire field range. After subtraction of the Oersted field, we find $B_{AD}=+1.17 \pm 0.01$~mT and $B_{FL}=+0.16 \pm0.08$~mT in Pt/Co, and $B_{AD}=-0.70\pm 0.01$~mT and $B_{FL}\leq-0.05\pm0.17$~mT in Ta/Co for $j=10^7$~A/cm$^{2}$. Note that, without taking into account the thermoelectric signal, the value of $B_{AD}$ in Ta/Co would be overestimated by 250-350\%, depending on the external field, as shown by the open triangles in Fig.~\ref{fig6} (b).

Table~\ref{table1} reports a summary of the results obtained in the present study. It is interesting to compare the SOTs measured here with those reported for thinner Pt/Co and Ta/Co layers with perpendicular magnetization, namely Pt(3nm)/Co(0.6)/AlO$_{\mathrm{x}}$ and Ta(3nm)/CoFeB(0.9nm)/MgO.~\cite{GarelloNN2013,AvciPRB2014} Comparison of torques in layers of different volume requires nomalization of SOT by the thickness of the magnetic layer. Once this is done, we find that $B_{AD}$ is comparable in the two sets of samples, whereas $B_{FL}$ is about an order of magnitude smaller in the thick FM relative to the thin ones. This result is not unexpected considering that $B_{FL}$ is associated either to a Rashba-like interface effect~\cite{MironNM2010,ManchonPRB2009} or to the field-like component of the spin Hall torque,~\cite{HaneyPRB2013} or a combination of both.~\cite{KimNM2013} Previous SOT measurements of perpendicularly magnetized Ta(1nm)/CoFeB(0.8-1.4nm)/MgO layers and in-plane magnetized Pt(3nm)/Co(1-3nm) also showed an increase of the $B_{AD}/B_{FL}$ ratio with increasing thickness of the FM.~\cite{KimNM2013,SkinnerAPL2014} Another way of comparing the AD SOT between different samples is to convert it into an effective spin Hall angle ($\theta_{SH}$), following a model in which the AD torque is entirely ascribed to the absorption of the spin current produced by the bulk spin Hall effect in the HM.~\cite{LiuS2012} Assuming a homogeneous current distribution within the bilayer (which sets a lower bound for the Ta/Co case) and and spin diffusion lengths $\lambda_{Pt}=1.4$~nm~\cite{LiuArx2011} and $\lambda_{Ta}=1.8$~nm,~\cite{HahnPRB2013} we obtain $\theta_{SH}= 0.144$ for Pt/Co and $\theta_{SH}=-0.086$ for Ta/Co, in agreement with Refs.~\onlinecite{GarelloNN2013,AvciPRB2014}.

\begin{table*}

\begin{tabular}{  p{5.2cm} | c | c | c | c | c  }
\hline \hline
@ $j=10^7$~A/cm$^2$ & $\rho$ ($\mu\Omega$cm) & $E_{\nabla T}$ (V/m) & $B_{AD}$ (mT) & $B_{FL+Oe}$ (mT) & $B_{Oe}$ (mT) (est.)\\
 \hline
Co(8nm) 									& 34.87 & +0.17 & $\approx$0 & $\approx$0  & $\approx$0   \\

Ti(6nm)/Co(2.5nm)            	   	 & 176.47 & +0.68 & $\approx$0 & -0.26$\pm$0.19 &  $\geq$-0.36  \\

Ta(1nm)/Cu(6nm)/Co(2.5nm)     & 17.42 & $\approx$0 & $\approx$0 & -0.22$\pm$0.06 &  $\leq$-0.36  \\

Ta(1nm)/Co(2.5nm)/Cu(6nm)     & 14.50 & +0.025 & $\approx$0 & +0.18$\pm$0.02 &  $\geq$+0.36  \\

Pt(6nm)/Co(2.5nm)     				  & 40.27 & $\approx$0 & +1.17$\pm$0.01 & -0.20$\pm$0.08 & $\approx$-0.36 \\

Ta(6nm)/Co(2.5nm)      				 & 142.94 & +1.06 & -0.70$\pm$0.01 & -0.41$\pm$0.17 & $\geq$-0.36   \\
\hline \hline

\end{tabular}

\caption{Summary of the results obtained in this work.} \label{table1}
\end{table*}

The thermoelectric contribution to the SOT measurements performed on perpendicular Ta(3nm)/CoFeB(0.9nm)/MgO was found to be less than 5$\%$,~\cite{AvciPRB2014} which is much smaller than that of the thicker Ta(6nm)/Co(2.5nm) bilayer studied here. This is due to two factors: first, for the same current density, the effect of the torque scales inversely with the thickness of the FM layer. Second, the ac susceptibility of the magnetization during a field sweep is larger in perpendicularly magnetized samples since external field, usually applied in-plane, pulls the magnetization away from the easy axis. Thus, the second harmonic SOT signal in the thin layers with out-of-plane easy axis is much larger than in relatively thick layers with in-plane magnetization. Accordingly, for $j = 10^7$~A/cm$^{2}$, we have reported $R^{2\omega}_{AD}\approx$15~m$\Omega$ in Ta(3nm)/CoFeB(0.9nm)/MgO, whereas $R^{2\omega}_{AD}\approx R^{2\omega}_{\nabla T}\approx$1~m$\Omega$ for Ta(6nm)/Co(2.5nm). With the same logic and taking into account that $R^{2\omega}_{\nabla T}\approx 0$ for Pt(6nm)/Co(2.5nm), thermoelectric contributions to the SOT measurements of thin perpendicular Pt/Co films are expected to be negligible.

\section{CONCLUSIONS}
\label{Conclusions}
In summary, we have presented a consistent method to separate SOT and thermoelectric effect measurements based on the harmonic analysis of the transverse resistance. The second harmonic transverse resistance $R^{2\omega}_{xy}$ consists of the sum of three components, $R^{2\omega}_{FL}+R^{2\omega}_{AD}+R^{2\omega}_{\nabla T}$, proportional, respectively, to the FL SOT, AD SOT, and vertical thermal gradient across the FM layer. Both $R^{2\omega}_{AD}$ and $R^{2\omega}_{\nabla T}$ have a $\cos\varphi$ dependence on the in-plane magnetization direction, which allows for the separation of these two components from $R^{2\omega}_{FL}$. Further separation of $R^{2\omega}_{AD}$ and $R^{2\omega}_{\nabla T}$ is possible by exploiting the field-dependence of the SOT-induced signal. Both macrospin simulations and measurements on a series of FM/LM and FM/HM bilayers validate this model. Although this paper is focused on in-plane magnetization systems, the model is also valid for systems with perpendicular magnetic anisotropy provided that the magnetization is tilted into the plane with an external field larger than the effective perpendicular anisotropy field.

Current injection in FM/NM bilayers creates perpendicular temperature gradients due to Joule heating and asymmetric heat dissipation towards the air and substrate side of the samples. Placing the less resistive layer on top and the more resistive layer on the bottom, next to the substrate, enhances the temperature gradient due to the larger Joule heating in the top layer and larger heat dissipation through the substrate. Inverting the position of the low and high resistivity layers results in a decrease or even the cancelation of the temperature gradient. Measurements of Ti/Co, Cu/Co, Co/Cu, and Co layers in which the ANE is the only contribution to the second harmonic transverse resistance agree with this picture.

In light of these results, we have studied Ta/Co and Pt/Co bilayers with large spin-orbit coupling. We found that the AD SOT is strong in both systems and comparable to that measured for thinner Pt/Co and Ta/CoFeB layers with perpendicular magnetization, once normalized by the thickness of the magnetic layer. The FL SOT is found to be about one order of magnitude smaller compared to the thin FM/NM layers. Additionally, we have found a significant thermoelectric signal in Ta/Co bilayers compatible with the sign of the ANE, which can lead to an overestimation of the AD SOT if not explicitly considered in the analysis of the second harmonic transverse voltage. Thermoelectric effects are found to be negligible for Pt/Co. By comparing the results obtained in this work with previous reports on thinner, perpendicularly magnetized bilayers, we find that thermoelectric effects in ac transverse resistance measurements become more influential in thick FM layers due to the relative decrease of the SOT signals with FM thickness. This scenario must be taken into account in thickness-dependent studies of SOTs.

Taken together, our results show that consistent measurements of SOTs and transverse thermoelectric effects can be performed in FM/NM systems, even when both provide non-negligible contributions to the ac transverse voltage. As FM/HM bilayers are of great interest for both the SOT and spin caloritronics fields, understanding the interplay of such phenomena may lead to a better control of the generation and detection of spin currents in these systems.

\begin{acknowledgments}
We thank Dr. Morgan Trassin for providing a test Pt/YIG sample for the measurements of the SSE in Pt. We acknowledge funding from the the Swiss National Science Foundation through Grant No. 200021-153404.
\end{acknowledgments}

\section{Appendix A: Harmonic analysis of the transverse voltage}
\label{AppendixA}
We perform here the harmonic analysis of the transverse voltage $V_{xy}(t)= R_{xy}(t)I_{0}\sin(\omega t)$, where $I_{0}\sin(\omega t)$ is the injected current and $R_{xy}(t)$ the transverse resistance, which takes into account also transverse thermoelectric effects.
To separate the dependence on static and dynamic parameters, we write the transverse resistance as $R_{xy}(t)=R_{xy}(\mathbf{B}_{0} + \mathbf{B}_{I}(t))$, where $\mathbf{B}_{0}$ represents the sum of the external and effective anisotropy fields and $\mathbf{B}_{I}=\mathbf{B}_{FL}+\mathbf{B}_{AD}+\mathbf{B}_{Oe}$ the sum of the current-induced fields including the Oersted term. In the limit of small oscillations of the magnetization, $R_{xy}(t)$ can be expanded to first order as
\begin{equation}
\label{eqRH}
R_{xy}(t)\approx R_{xy}(\mathbf{B}_{0}) + \frac{d R_{xy}}{d\mathbf{B}_{I}}\cdot \mathbf{B}_{I} \sin(\omega t) \, ,
\end{equation}
where $\mathbf{B}_{I}$ is the field produced by a current of amplitude $I_{0}$ and we assume a linear relationship between field and current. Inserting Eq.~\ref{eqRH} into the expression for the transverse voltage gives
\begin{equation}
\label{eqVH}
V_{xy}(t) \approx   I_{0}[R_{xy}^{0} + R_{xy}^{\omega}\sin(\omega t)+R_{xy}^{2\omega}\cos(2\omega t)] \, ,
\end{equation}
where $R_{xy}^{0}=\frac{1}{2}\frac{d R_{xy}}{d\mathbf{B}_{I}}\cdot \mathbf{B}_{I}$, $R_{xy}^{\omega}= R_{xy}(\mathbf{B}_{0})$, and $R_{xy}^{2\omega}=-\frac{1}{2}\frac{d R_{xy}}{d\mathbf{B}_{I}}\cdot \mathbf{B}_{I}$ are the zero, first, and second harmonic components of the transverse resistance, respectively. Note that $R_{xy}^{\omega}$ is equivalent to the transverse resistance of conventional dc measurements, whereas $R_{xy}^{2\omega}$ represents the modulation of
the transverse resistance due to the current-induced fields and thermoelectric effects. The first and second harmonic expressions for the transverse resistance can be written as
\begin{equation}
\label{eqRfApp}
R^{\omega}_{xy}=R_{AHE}\cos\theta+R_{PHE}\sin^{2}\theta\sin(2\varphi),
\end{equation}
\begin{multline}
\label{eqR2fApp}
R^{2\omega}_{xy}= (R_{AHE}-2R_{PHE}\cos\theta\sin(2\varphi))\frac{d\cos\theta}{d\mathbf{B}_{I}}\cdot \mathbf{B}_{I} \\
+R_{PHE}\sin^{2}\theta\frac{d\sin(2\varphi)}{d\mathbf{B}_{I}}\cdot \mathbf{B}_{I}+\alpha \nabla T I_{0} \sin\theta\cos\varphi],
\end{multline}
where($\theta,\varphi$) are the polar and azimuthal angles of the magnetization vector, respectively, as defined in Fig.~\ref{fig1} (a). To proceed further, the scalar products in Eq.~\ref{eqR2fApp} must be carried out by noting that the only component of the current-induced field that induces a change of the angle $\theta$ ($\varphi$) is the polar (azimuthal) one, which gives
\begin{eqnarray}
\label{eqDer}
\frac{d\cos\theta}{d\mathbf{B}_{I}}\cdot \mathbf{B}_{I} & = & \frac{d\cos\theta}{dB^{\theta}_{I}} B^{\theta}_{I}, \\ \frac{d\sin(2\varphi)}{d\mathbf{B}_{I}}\cdot \mathbf{B}_{I} & = & \frac{d\sin(2\varphi)}{dB^{\varphi}_{I}}B^{\varphi}_{I}.
\end{eqnarray}
The dependence of the magnetization angles on the current-induced field can be replaced by the dependence on the external field by substituting $dB^{\theta}_{I}$ with $dB^{\theta}_{ext}=B_{ext}d\sin(\theta_B - \theta_0)$ and $dB^{\varphi}_{I}$ with $dB^{\varphi}_{ext}=B_{ext}\sin\theta_B d\varphi$, where the external field is applied in the direction defined by ($\theta_B$, $\varphi$). Further, the derivatives with respect to the field that appear in Eqs.~\ref{eqDer} must be carried out with respect to the variable that is changed in the experiment. In previous work on SOTs we performed the harmonic transverse resistance analysis for field scans, in which the amplitude of $B_{ext}$ changes while its direction is fixed.~\cite{GarelloNN2013,AvciPRB2014} Here we analyze the complementary case of angle scans, where $B_{ext}$ is constant in amplitude and its direction changes. In such a case, Eq.~\ref{eqR2fApp} reads
\begin{multline}
\label{eqR2fangle}
R^{2\omega}_{xy}=(R_{AHE}-2R_{PHE}\cos\theta\sin(2\varphi))\frac{d\cos\theta}{d\theta} \frac{B^{\theta}_{I}}{-\cos(\theta_B-\theta)B_{ext}} \\
+R_{PHE}\sin^{2}\theta\frac{d\sin(2\varphi)}{d\varphi}\frac{B^{\varphi}_{I}}{\sin\theta_B B_{ext}}+\alpha \nabla T I_{0} \sin\theta\cos\varphi \, .
\end{multline}
If the external field is applied in-plane ($\theta_B=\pi/2$) and the samples have easy-plane anisotropy ($\theta \approx \pi/2$), as in the experiments presented Sect.~\ref{results}, Eq.~\ref{eqR2fangle} reads

\begin{equation}
\label{eqR2fsimpleApp1}
R^{2\omega}_{xy} = R_{AHE}\frac{d\cos\theta}{d\theta} \frac{B^{\theta}_{I}}{B_{ext}}
+R_{PHE}\frac{d\sin(2\varphi)}{d\varphi}\frac{B^{\varphi}_{I}}{B_{ext}}+\alpha \nabla T I_{0} \cos\varphi \, .
\end{equation}
By substituting $\dfrac{dR^{\omega}_{xy}}{d\theta}$ for $R_{AHE}\frac{d\cos\theta}{d\theta}$ and $\dfrac{dR^{\omega}_{xy}}{d\varphi}$ for $R_{PHE}\frac{d\sin(2\varphi)}{d\varphi}$ in Eq.~\ref{eqR2fsimpleApp1} we have

\begin{equation}
\label{eqR2fsimpleApp2}
R^{2\omega}_{xy} = \dfrac{dR^{\omega}_{xy}}{d\theta} \frac{B^{\theta}_{I}}{B_{ext}}
+\dfrac{dR^{\omega}_{xy}}{d\varphi} \frac{B^{\varphi}_{I}}{B_{ext}}+ I_{0} \alpha \nabla T \cos\varphi  \, .
\end{equation}
Since $\mathbf{B}_{AD} = B_{AD}(\mathbf{m}\times \mathbf{y})= B_{AD} \cos\varphi \, \mathbf{e}_{\theta}$ and $\mathbf{B}_{FL} = B_{AD}[\mathbf{m}\times(\mathbf{m}\times \mathbf{y})] = B_{AD} \cos\varphi \, \mathbf{e}_{\varphi}$, we have $B^{\theta}_{I} = B_{AD}\cos\varphi$ and $B^{\varphi}_{I} = B_{FL}\cos\varphi$. By substituting these expressions into Eq.~\ref{eqR2fsimpleApp2}, we finally obtain Eq.~\ref{eqR2fsimple} reported in Sect.~\ref{Harmonic analysis}:

\begin{equation}
\label{eqR2fsimpleApp3}
R^{2\omega}_{xy} = \dfrac{dR^{\omega}_{xy}}{d\theta} \frac{B_{AD}\cos\varphi}{B_{ext}}
+\dfrac{dR^{\omega}_{xy}}{d\varphi} \frac{B_{FL}\cos\varphi}{B_{ext}}+ I_{0} \alpha \nabla T \cos\varphi  \, .
\end{equation}

\bibliographystyle{apsrev4-1}

\bibliography{Manuscript_thermal_effects_vs_SOTs_submission_v1}

\end{document}